\documentclass[a4paper,prd,twocolumn,superscriptaddress,nofootinbib,showpacs,amsmath,amssymb]{revtex4}
\usepackage{graphicx,amsmath,amsfonts,amssymb,revsymb,dcolumn,epsfig,bm,hyperref,xspace}

\newcommand{\dct}{D}

\newcommand{\Rt}{\mathcal{R}}

\newcommand{\OO}[1]{\mathcal{O}\left(#1\right)}
\newcommand{\tc}{\prime}
\newcommand{\tctc}{{\prime\prime}}

\newcommand{\kp}{\kappa}
\newcommand{\Hc}{\mathcal{H}}
\newcommand{\zz}{z}

\newcommand{\EX}{\Theta}

\newcommand{\pa}{\hat{a}}
\newcommand{\dCIS}{\delta\sigma}

\newcommand{\dEX}{{\delta{}\EX}}
\newcommand{\dRt}{\delta{}\Rt}
\newcommand{\A}{\phi}
\newcommand{\AI}{\Phi}
\newcommand{\AIk}{\AI_k}
\newcommand{\CS}{\psi}
\newcommand{\CSI}{\Psi}
\newcommand{\E}{\mathcal{E}}
\newcommand{\B}{\mathcal{B}}
\newcommand{\dpp}{{\delta{}p{}}}
\newcommand{\ded}{{\delta\rho{}}}
\newcommand{\vus}{\mathcal{V}}
\newcommand{\dAPD}{\delta\Pi}
\newcommand{\dd}{\mathrm{d}}
\newcommand{\del}{\partial}
\newcommand{\MS}{\zeta}
\newcommand{\MSk}{\zeta_k}
\newcommand{\vp}{v_k}
\newcommand{\HR}{\mathrm{R}_H}
\newcommand{\wQ}{w_q}
\newcommand{\bbar}[1]{\bar{\bar{#1}}}
\newcommand{\bbbar}[1]{\bar{\bar{\bar{#1}}}}

\begin{document}
\title{Large Adiabatic Scalar Perturbations in a Regular Bouncing Universe}

\author{Sandro Dias Pinto Vitenti}
\email{vitenti@cbpf.br}
\author{Nelson Pinto-Neto}
\email{nelson.pinto@pq.cnpq.br}
\affiliation{Centro Brasileiro de Pesquisas F\'{\i}sicas,
Rua Dr.\ Xavier Sigaud 150 \\
22290-180, Rio de Janeiro -- RJ, Brasil}

\date{\today}

\begin{abstract}
It has been shown that a contracting universe with a dust-like ($w\approx0$) fluid may provide an almost scale invariant spectrum for the gravitational scalar perturbations. As the universe contracts, the amplitude of such perturbations are amplified. The gauge invariant variable $\AI$ develops a growing mode which becomes much larger than the constant one around the bounce phase. The constant mode has its amplitude fixed by Cosmic Background Explorer (COBE) normalization, thus the amplitude of the growing mode can become much larger than 1. In this paper, we first show that this is a general feature of bouncing models, since we expect that general relativity should be valid in all scales away from the bounce. However, in the Newtonian gauge, the variable $\AI$ gives the value of the metric perturbation $\A$, raising doubts on the validity of the linear perturbative regime at the bounce. In order to address this issue, we obtain a set of necessary conditions for the perturbative series to be valid along the whole history of the model, and we show that there is a gauge in which all these conditions are satisfied, for a set of models, if the constant mode is fixed by COBE normalization. As a by-product of this analysis, we point out that there are sets of solutions for the perturbation variables where some gauge-fixing conditions are not well defined, turning these gauges prohibited for those solutions.
\end{abstract}


\maketitle

\section{Introduction}

Cosmological models with a contracting phase preceding a bounce to our present expanding phase have been studied as extensions of the standard cosmological model. They were analyzed in several contexts, including regular and singular bounces~\cite{Gasperini1993, Gasperini1994, Wands1999, Brandenberger2001, Martin2002, Finelli2002, Lyth2002a, Lyth2002, Khoury2002, Durrer2002, Hwang2002, Cartier2003}. In~\cite{Wands1999,Finelli2002,Allen2004,Peter2007}, it was shown that if the contracting phase is dominated by a dust-like fluid and the perturbations are seeded by quantum vacuum fluctuations, the curvature perturbation $\MS$ develops an almost scale invariant spectrum.

There are various ways to obtain a bouncing cosmology. For each model there is a specific way to evolve the perturbations through the bounce phase. However, one can estimate the perturbations in the post-bounce phase by imposing some general continuity conditions on both background and perturbation variables. It was shown in~\cite{Finelli2002} that the curvature perturbation $\MS$ is amplified in the contracting phase, and remains constant and scale invariant in the expanding phase. Nevertheless, the Bardeen~\cite{Bardeen1980} potential $\AI$ develops a large growing mode in the contracting phase, which is converted entirely into a decaying mode in the expanding phase, differently from inflation, where the decaying mode is usually small. This behavior of the perturbations was also obtained using models for the bounce which allows complete calculations of the perturbations through the bounce~\cite{Peter2007}.

Even though the growing mode of $\AI$ couples only with the decaying mode in the expanding phase (see for example~\cite{Finelli2002}), its large value at the bounce raises a problem concerning the violation of linear perturbation theory around this time. In a previous work~\cite{Allen2004}, it was explicitly shown, using a specific model, that indeed the Bardeen potential grows larger than 1 at the bounce, but this mode can be made harmless with a suitable gauge choice. Thus remains the question about whether this problem is a general feature of bouncing models and, in case this is true, whether such gauge choices are still efficient to solve this problem in a broader class of models. In order to address this point, we show that the ratio between the Bardeen potential at the bounce and its constant value long after the bounce is indeed very large in general. Then we obtain a set of necessary conditions, for the metric and matter perturbations, that must be satisfied in a valid linear perturbation theory. The first part of this set is defined by imposing that the metric perturbations remain small when compared to their background values. The second one comes through the imposition that the perturbed Einstein equations remain small when compared with the background evolution. As these conditions are applied to the perturbations, it is necessary and sufficient that they should exists at least in one specific gauge. This means that it should have at least one way of embedding the reference metric in the spacetime in which the difference between the two metrics satisfies all conditions. Finally, we apply these conditions for the perturbations using a specific gauge, and we show that they are fully satisfied whenever the constant modes of $\AI$ or $\MS$ in the expanding phase are small.

This paper is organized as follows. In Sec.~\ref{sec:pert} we make a brief review of linear perturbation theory around a background with homogeneous and isotropic spatial sections. One can obtain an almost scale invariant spectrum when the perturbation freezes during a dustlike fluid domination. Therefore, in this work, we consider the models in which the spectrum of adiabatic perturbations is formed within this mechanism. In Sec.~\ref{sec:contracting}, we discuss the growth of adiabatic perturbations in the contracting phase of bouncing models with one or more fluids, and the relations between the gauge invariant curvature ($\MS$) perturbation and the Bardeen ($\AI$) potential. For the bounce crossing, we review in Sec.~\ref{sec:quantumbounce} a nonsingular bounce generated by quantum gravity effects in which these two gauge invariant variables remain finite and calculable. Then we show explicitly that the Bardeen potential becomes very large at the bounce. Next, in Sec.~\ref{sec:genbounce}, we discuss this issue for a class of models, and then compare with the results obtained in the quantum bounce scenario, arriving at the same conclusions. In Sec.~\ref{sec:pert:cond}, we obtain the set of necessary conditions that scalar adiabatic perturbations should satisfy in order to keep linear perturbation theory valid all along the contracting phase, during the bounce, and after it in the expanding phase before recombination. Then we show in Sec.~\ref{sec:pert:gauge}, using the uniform curvature gauge that all the above-mentioned conditions are satisfied along all these phases, except for the bounce itself, where in Sec.~\ref{sec:gauge:bounce} it is shown that one must use the synchronous gauge in order to keep all these necessary conditions satisfied. We end up with discussions and conclusions in Sec.~\ref{conclusions}.

\section{Linear cosmological perturbations}
\label{sec:pert}

From the Einstein-Hilbert action, one can obtain the second order Lagrangian for the Mukhanov--Sasaki~\cite{Mukhanov1992} variable
\begin{equation}\label{eq:L}
L_v = \int\dd^3x\frac{1}{2}\left(v^{\tc2}-c_s^2\delta^{ij}\del_iv\del_jv+\frac{z^{\tctc}}{z}v^2\right),
\end{equation}
yielding the equations of motion for their modes with wave number $k$,
\begin{equation}
\label{eq:muk}
\vp^{\tctc} + \left(c_s^2k^2-\frac{\zz^{\tctc}}{\zz}\right)\vp = 0,
\end{equation}
where
\begin{equation}\label{eq:zz}
\zz = \frac{\sqrt{\beta}}{x\mathcal{H}c_s}, \quad \beta = \frac{\kappa}{2}a^2\left(\rho+p\right), \quad c_s^2 = \frac{\dd p}{\dd\rho}.
\end{equation}
In this work, we define $\kappa = 8\pi{}G/c^4$, $\mathcal{H}\equiv a^{\tc}/a$, $a$ is the dimensionless scale factor of the background flat Friedmann model, a prime denotes derivative with respect to conformal time $\dd\eta=c\dd{}t/a$, $t$ being cosmic time, $x \equiv a_0/a$ is the red-shift function, $\rho$ and $p$ are the total energy density and pressure of the matter content of the model, respectively, and a subscript $0$ denotes the present value of the respective quantity; we define $\zz$ with an additional factor $a_0^{-1}$ compared to that defined in~\cite{Mukhanov1992}, Eq.~10.43b.

Defining $\Omega \equiv \rho/\rho_c$, where $\rho_c$ is the critical density today and using the energy conservation equation
\begin{equation}
\label{conservation}
\frac{\dd \rho}{\dd t} + 3H(\rho+p) = 0 \quad \rightarrow \quad
\frac{\dd\rho}{\dd x} = \frac{3\left(\rho + p\right)}{x},
\end{equation}
one obtains
\begin{align}\label{eq:z2}
\beta &= \frac{1}{2x \HR^2}\frac{\dd \Omega}{\dd x}, \quad \zz^2 = \frac{1}{2c_s^2x\Omega}\frac{\dd \Omega}{\dd x}, \\ \nonumber
c_s^2 &= \frac{x}{3}\left(\frac{1}{x^2}\frac{\dd \Omega}{\dd x}\right)^{-1}\frac{\dd}{\dd x}\left(\frac{1}{x^2}\frac{\dd \Omega}{\dd x}\right), \\
\zz^2 &= \frac{3}{2x^4\Omega}\left(\frac{\dd \Omega}{\dd x}\right)^2\left(\frac{\dd}{\dd x}\left(\frac{1}{x^2}\frac{\dd \Omega}{\dd x}\right)\right)^{-1},
\end{align}
where $\HR\equiv \Hc_0^{-1} = c/(a_0 H_0)$ is the comoving Hubble radius and $H_0$ is the present value of the Hubble function $H\equiv a^{-1}\dd{}a/\dd{}t$. Note also that $\mathcal{H} = E/(x \HR)$, where $E = H/H_0$. For a single fluid with $w = p/\rho$ constant, one gets $\Omega = \Omega_{w0} x^{3(1+w)}$ and
\begin{equation}\label{eq:single}
c_s^2 = w, \quad \zz = \sqrt{\frac{3(1+w)}{2w}}x^{-1}.
\end{equation}

From Eq.~\eqref{eq:single} we note that Eq.~\eqref{eq:muk} reduces to the equation for the perturbations found in~\cite{Pinho2007,Peter2007} for a single fluid dominated quantum bounce
\begin{equation}\label{eq:muk2}
\vp^{\tctc} + \left(w k^2-\frac{a^{\tc\tc}}{a}\right)\vp = 0,
\end{equation}
Therefore, we can use this same Eq.~\eqref{eq:muk} to evolve the perturbations in the classical contracting phase and through this particular quantum bounce~\cite{Lemos1996,Peter2007}, which we will present in the next section.

Equation~\eqref{eq:L} formally looks like a Lagrangian for a free scalar field with a time-dependent mass, which can be readily quantized. Whenever the potential $\zz^{\tc\tc}/\zz$ becomes negligible with respect to the wave number $k$, vacuum initial conditions can be imposed. This happens on the onset of inflation or in the far past in the contracting phase of bouncing models.

The Mukhanov--Sasaki variable has the following relation with the gauge invariant Bardeen potential and curvature perturbation~\cite{Mukhanov1992}, respectively,
\begin{equation}\label{eq:def:AI:MS}
D^2\AI = -\sqrt{4\pi}l_p{}x^2\Hc\zz^2\left(\frac{v}{\zz}\right)^{\tc},\quad \MS = -\sqrt{4\pi}l_p \frac{v}{\zz},
\end{equation}
where $l_p \equiv \sqrt{G\hbar/c^3}$ is the Planck length and $D^2$ the spatial Laplacian. Using Eq.~\eqref{eq:muk} one can also show that
\begin{equation}\label{eq:MS_AI}
\MS = \frac{1}{x^2c_s^2\zz^2}\left[\left(\frac{\AI}{\Hc}\right)^\tc + 2\AI\right].
\end{equation}
Since the Bardeen potential is a dimensionless quantity, $v$ must have dimensions of inverse length. From now on we will deal with dimensionless quantities by conveniently multiplying all physical quantities by appropriate powers of $\HR$ (e.g. $v \rightarrow v\HR$, $k \rightarrow k\HR$, $\eta \rightarrow \eta/\HR$, etc.).

From the operator decomposition,\footnote{The creation/annihilation operator decomposition of $\AI$ involves terms like $\AIk a_k^\dagger\dd^3k$. Since the operator $\AI$ is dimensionless and $a_k^\dagger$ has unity of length $L^{3/2}$, $\AIk$ also have dimension of length $L^{3/2}$.} we obtain
\begin{equation}\label{eq:Bpr}
k^2\AIk = \sqrt{4\pi}\frac{l_p}{\HR}x\sqrt{\Omega}z^2\left(\frac{v}{z}\right)^{\tc},
\end{equation}
Using the following definitions of power spectrum for the Bardeen potential and gauge invariant curvature perturbation, $\Delta_\AI^2 \equiv k^3\vert\AIk\vert^2 / (2\pi^2)$ and $\Delta_\MS^2 \equiv k^3\vert\MSk\vert^2 / (2\pi^2)$, respectively, we get
\begin{align}\label{eq:def:psphi}
\Delta_\AI &= \sqrt{\frac{2}{\pi}}\frac{l_p}{\HR}\frac{x\sqrt{\Omega}}{k^2}k^{3/2}z^2
\left\vert\left(\frac{v}{z}\right)^{\tc}\right\vert, \\ \nonumber
\Delta_\MS &= \sqrt{\frac{2}{\pi}}\frac{l_p}{\HR}k^{3/2}\left\vert\frac{v}{z}\right\vert
\end{align}
Note that the amplitude of the perturbations is multiplied by a very small number $l_p/\HR = 9.9\times 10^{-62}h$, where $h = H_0 / (100\;\mathrm{Km\,s^{-1}Mpc^{-1}})$ is the dimensionless Hubble constant.

The general solution of the mode Eq.~\eqref{eq:muk} can be expanded in powers of $k^2$ according to the formal solution~\cite{Mukhanov1992}
\begin{equation}\label{solform}
\begin{split}
\frac{v}{\zz} =  A_1(k)\left[1 - k^2 \int^\eta_0\frac{\dd\bar{\eta}}{\bar{\zz}^2} \int^{\bar{\eta}}_0\dd\bbar{\eta} \bbar{c}_s^2 \bbar{\zz}^2 + \dots\right] \\
+ A_2(k) \left[\int^\eta_{\eta^*}\frac{\dd\bar{\eta}}{\bar{\zz}^2} - k^2 \int^\eta_0\frac{\dd\bar{\eta}}{\bar{\zz}^2} \int^{\bar{\eta}}_0\dd\bbar{\eta} \bbar{c}_s^2 \bbar{\zz}^2\int^{\bbar{\eta}}_{\eta^*} \frac{\dd\bbbar{\eta}}{\bbbar{\zz}^2} + \dots\right],
\end{split}
\end{equation}
where we have shown the terms up to order $\OO{k^{2}}$, $\eta^{*}$ is a convenient choice of the integration constant, related to the conformal time where the initial conditions are set (in the case of bouncing models we make the choice $\eta^{*}\rightarrow-\infty$). The bounce takes place at $\eta = 0$ and any function with an over bar refers to its value calculated at $\bar{\eta}$, e.g., $\bar{f} = f(\bar{\eta})$. In Eq.~\eqref{solform}, the coefficients $A_1$ and $A_2$ are two constants depending only on the wave-number $k$ through the initial conditions. Once the solutions freeze ($c_s^2 k^2\ll z^\tctc/z$), i.e., when the mode is below its potential, the superhorizon solutions above can be used. Since we are interested in what happens with the amplitude of the spectrum after it is formed in a dust-dominated evolution, it is enough to analyze the superhorizon solutions. For long wavelengths of cosmological relevance, this happens during the bounce, of course, and around our expanding epoch.
Under these conditions, we can use Eq.~\eqref{solform} up to second order in $k^2$ to calculate $(v/\zz)^\tc$ in these situations, which reads
\begin{equation}\label{eq:psphi}
\begin{split}
\zz^2\left(\frac{v}{\zz}\right)^\tc &\simeq -k^2A_1(k)\int^{\eta}_0\dd\bar{\eta} \bar{c}_s^2\bar{\zz}^2 \\
&+ A_2(k) \left(1 - k^2\int_0^\eta\dd\bar{\eta} \bar{c}_s^2 \bar{z}^2  \int^{\bar{\eta}}_{-\infty}\frac{\dd\bbar{\eta}}{\bbar{\zz}^2}\right).
\end{split}
\end{equation}

\section{The problem}
\label{sec:contracting}

In this section, we will calculate $\Delta_\AI$ for long wavelengths using Eq.~\eqref{eq:psphi} after the decaying mode becomes negligible in the expanding phase $\Delta_\AI^0$ and near the bounce $\Delta_\AI^b$. We will show exactly for the quantum bounce of Refs.~\cite{Alvarenga2002,Peter2007}, and using general arguments for general bounces that $\Delta_\AI^b$ is many orders of magnitude larger than $\Delta_\AI^0$. Nevertheless, in the next section, we will show how the linear perturbation theory is still reliable near the bounce in spite of the problem, which will be described in the sequel.

Let us evaluate $\Delta_\AI^b$ and $\Delta_\AI^0$. In the case of $\Delta_\AI^b$, we can see from Eq.~\eqref{eq:psphi} that the term multiplying $A_1(k)$ is the decaying mode of the contracting phase, which goes to zero at the bounce. Hence, when the solution gets close enough to the bounce, the important contribution for $\Delta_\AI^b$ is
\begin{equation}\label{psphib}
\Delta_\AI^b = \sqrt{\frac{2}{\pi}}\frac{l_p}{\HR}\frac{x_b\sqrt{\Omega_b}}{k^2}k^{3/2}
\left\vert A_2(k)\right\vert.
\end{equation}
For $\Delta_\AI ^0$, the term multiplying $A_1(k)$ is the growing mode of the expanding phase. However, the last term which multiplies $A_2(k)$ can be written as
\begin{equation}\label{lastterm}
\begin{split}
&k^2\int_0^\eta\dd\bar{\eta} \bar{c}_s^2 \bar{\zz}^2\int^{\bar{\eta}}_{-\infty}\frac{\dd\bbar{\eta}}{\bbar{\zz}^2} \\
&= k^2\int_0^\eta\dd\bar{\eta} \bar{c}_s^2\bar{\zz}^2 \left(
\int^{\infty}_{-\infty}\frac{\dd\bbar{\eta}}{\bbar{\zz}^2}-
\int^{\infty}_{\bar{\eta}}\frac{\dd\bbar{\eta}}{\bbar{\zz}^2}\right),  \\
&\approx k^2 B \int_0^\eta\dd\bar{\eta}\bar{c}_s^2\bar{\zz}^2,
\end{split}
\end{equation}
where we defined the constant $B \equiv \int^{\infty}_{-\infty}\dd\bbar{\eta}\bbar{\zz}^{-2}$. We also have discarded the last term in the sum because it corresponds to a decaying mode in the expanding phase, and by the positivity of the integrand, it will always be smaller than $B$.

The presence of $B$, as we will see, makes the term multiplying $A_2(k)$ much more important than the one multiplying $A_1(k)$ in the evaluation of $\Delta_\AI^0$. Hence, we get
\begin{equation}\label{psphi0}
\Delta_\AI^0 = \sqrt{\frac{2}{\pi}}\frac{l_p}{\HR}x\sqrt{\Omega}
k^{3/2} \left\vert A_2(k) B \int_0^\eta\dd\bar{\eta}\bar{c}_s^2\bar{\zz}^2\right\vert,
\end{equation}
when the decaying mode becomes negligible.

The variable $\MS$ has a much simpler evolution, in the contracting phase, the term $$A_2(k) \int^\eta_{-\infty}\frac{\dd\bar{\eta}}{\bar{\zz}^2} = A_2(k)\left(B - \int^{\infty}_{\bar{\eta}}\frac{\dd\bbar{\eta}}{\bbar{\zz}^2}\right),$$ grows as the perturbations approach the bounce. In the expanding phase the integral can be split as in Eq.~\eqref{lastterm}, where the first term is constant and the second becomes the decaying mode. Unlike $\AI$, this decaying mode around the bounce has the same order of magnitude than its constant mode. Thus, after the bounce the power spectrum of $\MS$ is 
\begin{equation}\label{eq:DMS}
\Delta_\MS^0 = \sqrt{\frac{2}{\pi}}\frac{l_p}{\HR}k^{3/2}\left\vert A_2(k)B\right\vert.
\end{equation}
Generally, as we will show in the next sections, $$x\sqrt{\Omega}\int_0^\eta\dd\bar{\eta}\bar{c}_s^2\bar{\zz}^2 = \OO{1}.$$ Therefore, using Eqs.~\eqref{psphi0} and \eqref{eq:DMS}, we note that in the expanding phase $\Delta_\MS^0 \approx \Delta_\AI^0$. This shows that the power spectrum of $\MS$ is enough to assess the spectrum of $\AI$ in what concerns its constant mode. However, it is insensitive to the large growing/decaying mode developed by $\AI$.

We will now evaluate $\Delta_\AI^b$ and $\Delta_\AI^0$ for a particular quantum bounce \cite{Alvarenga2002,Pinto-Neto2005,Peter2007} and then for a more general case.

\subsection{The example of the quantum bounce}
\label{sec:quantumbounce}

References~\cite{Alvarenga2002,Pinto-Neto2005,Peter2007} show the canonical quantization of a minisuperspace cosmological model describing a perfect fluid with $p=\wQ\rho$ on a Friedmann geometry with flat spacelike hyper-surfaces.
The corresponding Wheeler-DeWitt equation reads
\begin{equation}
i\frac{\del\Psi_{(0)}(a,T)}{\del T}= \frac{1}{4}
\frac{\del^2\Psi_{(0)}(a,T)}{\del \chi^2}, \label{es202}
\end{equation}
where $\chi \equiv a^{3(1-\wQ)/2}2/(3(1-\wQ))$. This is just the time-reversed Schr\"odinger equation for a one-dimensional free particle constrained to the positive axis.

\begin{widetext}
After imposing a Gaussian initial wave function, the wave solution of this equation for all times in terms of $a$ reads
\begin{equation}\label{psi1t}
\Psi_{(0)}(a,T)=\left[\frac{8 T_b}{\pi\left(T^2+T_b^2\right)}
\right]^{\frac{1}{4}}
\exp\left[\frac{-4T_ba^{3(1-\wQ)}}{9(T^2+T_b^2)(1-\wQ)^2}-i\left[\frac{4Ta^{3(1-\wQ)}}{9(T^2+T_b^2)(1-\wQ)^2}
+\frac{1}{2}\arctan\biggl(\frac{T_b}{T}\biggr)-\frac{\pi}{4}\right]\right],
\end{equation}
where $T$ is related to conformal time through $\dd\eta = \left[a(T)\right]^{3\wQ-1} \dd T$ and $T_b$ is an arbitrary constant related to the width of the initial Gaussian.
\end{widetext}

Because of the chosen factor ordering, the probability density $\rho_\Psi(a,T)$ has a non trivial measure and it is given by
$\rho_\Psi(a,T)=a^{(1-3\wQ)/2}\left|\Psi_{(0)}(a,T)\right|^2$. Its continuity equation coming from Eq.~\eqref{es202} reads
\begin{equation}\label{cont}
\frac{\del\rho_\Psi}{\del T}
-\frac{\del}{\del a}\left[\frac{a^{(3\wQ-1)}}{2}
\frac{\del S}{\del a}\rho_\Psi\right] = 0,
\end{equation}
where $S$ denotes the imaginary phase of the wave function $\Psi_{(0)}$. This implies in the de Broglie--Bohm interpretation \cite{Bohm1952,Bohm1952a,Bohm1993,Holland1993} that
\begin{equation}\label{guidance}
\frac{\del a}{\del T}=-\frac{a^{(3\wQ-1)}}{2} \frac{\del{}S}{\del a},
\end{equation}
in accordance with the classical relations $\del{}a/\del{}T=\{a,H\}= -a^{(3\wQ-1)}P_a/2$ and $P_a=\del S/\del a$.

Inserting the phase of Eq.~\eqref{psi1t} into Eq.~\eqref{guidance}, we obtain the Bohmian quantum trajectory for the scale factor:
\begin{equation}
\label{at} a(T) = a_b
\left[1+\left(\frac{T}{T_b}\right)^2\right]^{1/(3(1-\wQ))}.
\end{equation}
Note that this solution has no singularities and tends to the classical solution when $T\rightarrow\pm\infty$.
Solution \eqref{at} can be obtained for other initial wave functions (see Ref.~\cite{Alvarenga2002}).

Changing to cosmic time $\dd t = a^{3\wQ}\dd T$, we obtain the Hubble function
\begin{equation*}
\frac{1}{a}\frac{\dd a}{\dd t} = H(t) = \frac{2Ta^{-3\wQ}}{3\left(T^2+T_b^2\right)\left(1-\wQ\right)}.
\end{equation*}
Solving Eq.~\eqref{at} for $T$,
\begin{equation*}
T = \pm T_b\sqrt{\left(\frac{a}{a_b}\right)^{3(1-\wQ)}-1},
\end{equation*}
we obtain
\begin{equation}\label{eq:QH2}
H^2 = \frac{4a_0^{-6\wQ}x_b^{-3(1-\wQ)}}{9\left(1-\wQ\right)^2T_b^2}\left({x^{3(1+\wQ)}-\frac{x^6}{x_b^{3(1-\wQ)}}}\right),
\end{equation}
where $x_b\equiv a_0/a_b$.

Equation~\eqref{eq:QH2} is equivalent to the Friedmann equation
\begin{equation}\label{eq:QH22}
H^2 = \frac{\kappa{}c^2}{3}\rho - H_0^2\Omega_{q0} x^6,
\end{equation}
with the additional term $-H_0^2\Omega_{q0} x^6$. The quantity $\Omega_q = \Omega_{q0} x^6$ is an effective density related to the quantum evolution. Thus, the evolution of the scale factor is equivalent to the one obtained by adding to the matter content, in the classical Friedmann equation, a stiff negative energy fluid, i.e., $\rho \rightarrow \rho + \rho_s$, where $\rho_s = -\Omega_{q0}\rho_c x^{3(1+\wQ)}$. Note, however, that this is a quantum effect and therefore there is no perturbation associated to this effective fluid. Note also that away from the bounce phase ($H_0^2\Omega_{q0} x^6 \ll \kappa{}c^2\rho/3$) we obtain $H^2 = H^2_0\Omega_{w0}x^{3(1+\wQ)}$, where
\begin{equation}
\Omega_{w0} = \left[\frac{2 a_0^{-3\wQ}x_b^{-3(1-\wQ)/2}}{3(1-\wQ)T_b}\right]^2.
\end{equation}

Comparing Eqs.~\eqref{eq:QH2} and \eqref{eq:QH22}, we obtain the important relation for the dimensionless $T_b$,
\begin{equation}
\label{eq:tb}
T_b = \frac{2 a_0^{1-3\wQ}x_b^{-3(1-\wQ)/2}}{3(1-\wQ)\sqrt{\Omega_{w0}}}.
\end{equation}

In this bouncing model there is one single fluid. Hence, Eq.~\eqref{eq:single} holds. Using Eqs.~\eqref{at} and \eqref{eq:tb}, we obtain
\begin{equation}
\label{B}
B = \int^{\infty}_{-\infty}\frac{\dd\eta}{\zz^2} = \frac{4\wQ\pi}{9(1-\wQ^2)\sqrt{\Omega_{w0}}}x_b^{3(1-\wQ)/2},
\end{equation}
and
\begin{equation}
\label{I2}
\int_0^\eta c_s^2 \zz^2 \dd\bar{\eta} \approx \frac{2}{5+3\wQ}\frac{3(1+\wQ)}{2x\sqrt{\Omega_w}},
\end{equation}
where $\Omega_w = \Omega_{0w} x^{3(1+\wQ)}$; in the last equation we have neglected the evaluation of the primitive at $\eta =0$. Note that in this simple model one must have $\wQ \approx 0$ in order to obtain an almost scale invariant spectrum of perturbations. Hence, as $x_b=a_0/a_b \gg 10^{10}$ (we expect that the bounce occurs much earlier then nucleosynthesis), the quantity $B$ is indeed a large number.

From Eqs.~\eqref{psphib} and \eqref{psphi0} we obtain
\begin{equation}\label{psphibq}
\Delta_\AI^b = \sqrt{\frac{2}{\pi}}\frac{l_p \sqrt{\Omega_{0w}}}{\sqrt{k}\HR}
\left\vert A_2(k)\right\vert x_b^{(5+3\wQ)/2},
\end{equation}
and
\begin{equation}\label{psphi0q}
\Delta_\AI^0 = \frac{4 l_p\sqrt{2\pi} \wQ k^{3/2}\left\vert A_2(k)\right\vert}{\HR\sqrt{\Omega_{0w}}3(1-\wQ)(5+3\wQ)}
 x_b^{3(1-\wQ)/2}.
\end{equation}
As expected, $\Delta_\AI^0$ does not depend on time. The ratio between these two
quantities is
\begin{equation}\label{ratio}
\frac{\Delta_\AI^b}{\Delta_\AI^0} = \frac{\left\vert\AIk^b\right\vert}{\left\vert\AIk^0\right\vert} = \frac{\Omega_{0w}3(1-\wQ)(5+3\wQ)}{4\pi(\sqrt{\wQ}k)^2} x_b^{1+3\wQ}.
\end{equation}
As the Cosmic Microwave Background CMB observations~\cite{Komatsu2011} require $\Delta_\AI^0 \approx 10^{-5}$ and $x_b \gg 10^{10}$, then $\Delta_\AI^b \gg 1$, which turns questionable the validity of linear perturbation theory at the bounce. We will now see that this issue is also present in a much larger class of bouncing models. 

\subsection{More general bounces}
\label{sec:genbounce}

The general solution for the Mukhanov--Sasaki variable, Eq.~\eqref{solform}, for the adiabatic perturbations, is valid in the contracting and expanding phases when the dynamics are given by general relativity (GR) and also through the bounce itself, in the case of the quantum bounce discussed above. For general bounces, one is not sure that the solution in Eq.~\eqref{solform} is valid through the bounce due to not having any particular analytic solution in order to evaluate it away from the bounce, as we did in the last subsection. However, if the bounce is short enough an estimate of Eq.~\eqref{solform} away from the bounce, when GR is valid, will be sufficient to evaluate the orders of magnitude of Eqs.~\eqref{psphib} and \eqref{psphi0} as long as a short bounce does not change these figures too much due to the expected continuity of perturbations through it (see~\cite{Deruelle1995} for a general discussion about matching conditions using the continuity of the perturbations and~\cite{Finelli2002} for its use in this context). However, one could have a bouncing model with a long-time scale. Thus, in our analysis we are assuming that the characteristic time of the bouncing model is small enough (usually of the order of $l_p$) that we can ignore this phase.

In order to estimate \eqref{psphib} and \eqref{psphi0} we must evaluate $$B \equiv \int^{\infty}_{-\infty}\frac{\dd\eta}{\zz^{2}}\quad\text{and}\quad I\equiv \int_0^\eta\dd\bar{\eta}\bar{c}_s^2\bar{\zz}^2.$$

For $B$, we first divide the integral in the pre- and post-bounce branches
\begin{equation}
B = \int^\infty_{-\infty}\frac{\dd\bar{\eta}}{\bar{\zz}^2} = \int^0_{-\infty}\frac{\dd\bar{\eta}}{\bar{\zz}^2} + \int^\infty_{0}\frac{\dd\bar{\eta}}{\bar{\zz}^2}.
\end{equation}
In each branch the scale factor and, consequently, the red-shift variable $x$ can be used as a time variable. Performing this transformation, we have
\begin{eqnarray}
\int^0_{-\infty}\frac{\dd\eta}{\zz^2} &=& -\int^{x_b}_0\frac{\dd{}x}{E^-\zz^2} = \int^{x_b}_0\frac{\dd{}x}{\vert{}E^-\vert\zz^2}, \\
\int^\infty_{0}\frac{\dd\eta}{\zz^2} &=& -\int^{0}_{x_b}\frac{\dd x}{E^+\zz^2} = \int^{x_b}_0\frac{\dd x}{E^+\zz^2},
\end{eqnarray}
where $E^-$ is the dimensionless Hubble function $H/H_0$ during the contracting phase and, therefore, a negative quantity; $E^+$ is the dimensionless Hubble function during the expanding phase. We have assumed that $x(\pm\infty) = 0$ but this is not necessarily true. However, one can always assume that $x(\pm\infty)/x_b \ll 1$, which is sufficient to estimate the integral. Also, an asymmetric bounce will not, in general, change too much the orders of magnitude we will evaluate. Hence, we will assume for simplicity that the bounce is symmetric
\begin{equation}
\int^{x_b}_0\frac{\dd x}{\vert E^-\vert\zz^2} = \int^{x_b}_0\frac{\dd x}{E^+\zz^2} \equiv \int^{x_b}_0\frac{\dd x}{E\zz^2}.
\end{equation}

Now we divide the above integral in two domains,
\begin{equation}
\int^{x_b}_0\frac{\dd x}{E\zz^2} = \int^{x_c}_0\frac{\dd x}{E\zz^2} + \int^{x_b}_{x_c}\frac{\dd x}{E\zz^2},
\end{equation}
where $x_c$ is the value of the red-shift function where the new physics of the bounce begin to be relevant ($x_c$ could be defined as the value of $x$ in which $\dd^2 a/\dd{}t^2 = 0$, the transition from the decelerating behavior typical of GR to the accelerating phase of the bounce. Note that the second portion of the sum above cannot be written if the solution \eqref{solform} is not valid through the bounce. However, as $x_c$ is generally of the same order of magnitude as $x_b$, then $x_c \gg 1$ (see the quantum bounce example, where $x_c = [(1+3\wQ)/4]^{1/[3(1-\wQ)]} x_b \approx x_b$ for $0<\wQ<1$). In this case, the interval $x_c < x < x_b$ is irrelevant when compared to $0 < x < x_c$ and, therefore,
\begin{equation}
\label{intgen1}
\int^{x_b}_0\frac{\dd x}{E\zz^2} \approx \int^{x_c}_0\frac{\dd x}{E\zz^2}.
\end{equation}


In the GR domain, we can use Eqs.~\eqref{conservation} and \eqref{eq:z2} to obtain
\begin{equation}
\label{approxz2}
\frac{x}{E\zz^2} = \frac{2c_s^2 x^3}{3E(1+p/\rho)}.
\end{equation}
As $x$ increases, the quantities above become dominated by the fluid with largest value of $p_q/\rho_q = \wQ$, and in this phase the integrand is dominated by the term $x^{3(1-\wQ)/2}$ since in one fluid domination $c_s^2$ and $p/\rho$ become constant and $E \propto x^{3(1+\wQ)/2}$. For simplicity, we are assuming that the fluid which dominates in this epoch has constant equation of state. As $x$ varies several orders of magnitude in the integration interval, the value of the integral is dominated by the integrand near instant $x_c$, where we are assuming that the fluid with equation of state $\wQ$ dominates. Using this feature, we show in the Appendix that this integral is approximated by Eq.~\eqref{eq:app:intfapprox}, i.e.,
\begin{equation}\label{eq:intfapprox}
\int^{x_c}_0\frac{\dd{}x}{E\zz^2} \approx \frac{2}{3(1-\wQ)}\frac{x_c}{E(x_c)\zz(x_c)^2}.
\end{equation}

As $x_c \approx x_b$, we obtain
\begin{equation}\label{Bapprox}
B \approx \frac{2}{3(1-\wQ)}\frac{2x_b}{E(x_b)\zz^2(x_b)},
\end{equation}
where it must be understood that, although evaluated at $x_b$, the functions $E(x)=\sqrt{\Omega(x)}$ and $\zz^2(x)$ in Eq.~\eqref{Bapprox} are the usual GR expressions for them, which are valid just before the bounce.

Note that if the integral $\int^{x_b}_{x_c} \dd{}x / (E\zz^2)$ makes sense during the bounce, it is a positive quantity that is being neglected in the evaluation of $B$ and, hence, Eq.~\eqref{eq:intfapprox} continues to be valid. Note also that although $E=0$ at the bounce, this integral converges for regular bounces (see the quantum bounce above).

For the second integral one has
\begin{align}
I &= \int^\eta_0\dd\bar{\eta}\bar{c}_s^2\bar{\zz}^2 = \int^a_{a_b}\dd\bar{a}\frac{a_0\bar{c}_s^2\bar{\zz}^2}{\bar{a}^2\bar{E}}, \nonumber \\
&\approx \int^a_{a_c}\dd\bar{a} \frac{a_0\bar{c}_s^2\bar{\zz}^2}{\bar{a}^2 \bar{E}} =
\int^a_{a_c}\dd\bar{a}\frac{3(1+\bar{p}/\bar{\rho})}{2 a_0 \bar{E}},
\end{align}
where, as justified before, we are taking $a_c \approx a_b$, and we have used Eq.~(\ref{approxz2}) 
for the last equality. The integrand in the last integral is an increasing function of $\bar{a}$, 
hence, using the mean value theorem, we get
\begin{equation}
\label{Iapprox}
I \approx  \int^a_{a_c}\dd\bar{a} \frac{a_0\bar{c}_s^2\bar{\zz}^2}{\bar{a}^2 \bar{E}} = \left.\frac{a_0c_s^2\zz^2}{a^2 E}\right\vert_{a_\star}(a-a_c) \lesssim \frac{c_s^2\zz^2 x}{E},
\end{equation}
where in the last approximation we used $a \gg a_c$ and $a_c \leq a_\star \leq a$.

Now inserting Eqs.~\eqref{Bapprox} and \eqref{Iapprox} into Eq.~\eqref{psphi0}, assuming for simplicity that near the bounce there is domination of one fluid, we get
\begin{equation}\label{psphi0gen}
\Delta_\AI^0 \lesssim \sqrt{\frac{2}{\pi}}\frac{4l_p \wQ k^{3/2}\left\vert A_2(k)\right\vert }{3(1-\wQ)\HR\sqrt{\Omega_{w0}}}
x_b^{3(1-\wQ)/2},
\end{equation}
which, apart from numerical factors of order unity, coincides with Eq.~\eqref{psphi0q}.

For the general $\Delta_\AI^b$ one has
\begin{equation}\label{psphibgen}
\Delta_\AI^b = \sqrt{\frac{2}{\pi}}\frac{l_p \sqrt{\Omega_{w0}}}{\sqrt{k}\HR}
\left\vert A_2(k)\right\vert x_b^{(5+3\wQ)/2},
\end{equation}
and the ratio is
\begin{equation}\label{ratiogen}
\frac{\Delta_\AI^b}{\Delta_\AI^0} = \frac{\left\vert\AIk^b\right\vert}{\left\vert\AIk^0\right\vert} \gtrsim \frac{\Omega_{0w}3(1-\wQ)}{4(\sqrt{\wQ}k)^2} x_b^{1+3\wQ}.
\end{equation}

Again, this ratio is proportional to $x_b^{(1+3\wQ)}$, and for any fluid with $\wQ>-1/3$ (assuming COBE normalization) one gets a very large amplitude during the bounce. As we have shown, this large ratio $\Delta_\AI^b/\Delta_\AI^0$ is a general feature of a long contracting phase and, therefore, it will be present in any reasonable bouncing model where the matter content satisfies $-1/3 < p/\rho < 1$.

\section{The solution}
\label{sec:pert:cond:gauge}

As we have seen, the gauge invariant Bardeen potential $\AI$ may grow in the bounce because what would be the decaying mode in the expanding phase is the growing mode in the contracting phase. This mode can be very large around the bounce if the contraction is huge. Then one could put into question the validity of linear perturbation theory at the bounce, which compromises all calculations of cosmological perturbations in bouncing models. However, the definition of a gauge invariant quantity is not unique. One simple reason for this is that one can multiply any gauge invariant quantity by a background function and it continues to be a gauge invariant quantity. For instance, if one defines the gauge invariant function $(a/a_0)^\alpha \Phi$ with $\alpha$ positive, which coincides with the Bardeen potential today, it is trivial to find a power $\alpha$, where this gauge invariant function is small at the bounce. Hence, what one has to do is to look at Einstein's equations for the perturbations themselves and see if the linear theory makes sense, at least in some gauge, during the whole history of the model before the usual epoch where nonlinearities become important. Note that it is not necessary that the theory makes sense in all gauges; a valid gauge transformation which relate different gauge choices at some phase in the cosmological evolution may not exist.

\subsection{Conditions for linearity}
\label{sec:pert:cond}

Let us concentrate on the scalar perturbations. The geometry of spacetime is given by
\begin{equation}
\label{perturb}
g_{\mu\nu}=g^{(0)}_{\mu\nu}+h_{\mu\nu},
\end{equation}
where $g^{(0)}_{\mu\nu}$ represents the homogeneous and isotropic cosmological background
\begin{equation}
\label{linha-fried}
\dd s^{2}=g^{(0)}_{\mu\nu}\dd x^{\mu}\dd x^{\nu}=-c^2\dd{}t^2 + a^{2}\gamma_{ij}\dd x^{i}\dd x^{j},
\end{equation}
where $\gamma_{ij}$ is the metric of the maximally symmetric spatial hyper-surfaces with normalized scalar curvature $K=0,\pm 1$, and $h_{\mu\nu}$ represents linear scalar perturbations around it, which we decompose into
\begin{align}
\label{perturb-componentes}
h_{00} &= 2\phi \nonumber, \\
h_{0i} &= -aD_i \B, \\
h_{ij} &= 2a^{2}(\psi\gamma_{ij}-D_i D_j\E), \nonumber
\end{align}
where $D_i$ is the covariant derivative with respect to $\gamma_{ij}$. Hence, the first conditions are
\begin{equation}\label{cond1}
\phi \ll 1,\; \B \ll 1,\; \E \ll 1,\; \psi \ll 1.
\end{equation}

Computing the perturbed Einstein equations in the background Gaussian coordinate system, one gets
\begin{align}\label{eq:def:friedpA}
\delta G_0{}^0 &= -\frac{2\EX\dEX}{3} -\frac{\dRt}{2} = -\kp \ded = \kp\delta T_0{}^0, \\ \label{eq:def:friedpB}
\delta G_i{}^0 &= \frac{2}{3} D_i\left(\dEX - \frac{3K}{a^2}\dCIS - \frac{\dct^2}{a^2}\dCIS\right), \\ \nonumber
&= D_i[\kp(\rho + p)\vus] = \kp\delta T_i{}^0\\ \label{eq:def:friedpCST}
\delta G_i{}^i &= -2\left(\frac{\dEX^\tc}{a}+\EX\dEX+\A\frac{\EX^\tc}{a} - \dct_i\pa^i + \frac{\dRt}{4}\right),\\ \nonumber
&= 3\kp\dpp = \kp\delta T_i{}^i, \\ \label{eq:def:friedpCSD}
\delta G_j{}^i &= - \frac{D^i D_j}{a^2}\left(\CS - \A - \frac{\dCIS^\tc}{a} - \frac{\EX}{3}\dCIS\right), \\ \nonumber
&= - \kp \frac{D^i D_j}{a^2}\dAPD = \kp \delta T_j{}^i,
\end{align}
where in the fourth equation $i\neq j$, $\dct^2\equiv D_iD^i$, $\ded$, and $\dpp$ are the perturbed energy density and pressure, respectively; $\dAPD$ is the anisotropic pressure, which we will consider to be null, and $\vus$ is the perturbed velocity field potential.

In these equations, the quantities $\dEX$, $\dCIS$, $\dRt$ and $a^i$ are the perturbed expansion rate, shear, curvature scalar, and worldline acceleration with respect to the constant cosmic time hyper-surfaces. The background expansion rate is simply $\EX = n^{\mu}{}_{;\mu} = 3H/c = 3\Hc/a$, where $n^{\mu}$ is the normal of the maximally symmetric spacelike hyper-surfaces and $_;$ represent the covariant derivative compatible with the background metric.

These quantities are related to the metric perturbations by
\begin{align}\label{dsigma}
\dCIS &= -a(\E^\tc-\B), \\ \label{dtheta}
a\dEX &= - \dct^2(\E^\tc-\B)+3\left(\Hc\phi+\psi^\tc\right), \\ \label{dR}
\dRt &= -\frac{4}{a^2}\left(\dct^2+3K\right)\psi, \\\label{da}
a^i &= -\frac{D^i\phi}{a^2}.
\end{align}
The gauge invariant variables are defined by the following combinations of the perturbation variables:
\begin{equation}
\AI = \A + \frac{\dCIS^\tc}{a}, \quad \CSI = \CS - \frac{\Hc\dCIS}{a}.
\end{equation}

We have to verify whether the perturbed Einstein equations remain small when compared with the background Einstein equations, where the non-null background Einstein's tensor components are
\begin{align}
G_0{}^0 &= -\left(\frac{\EX^2}{3}+\frac{3K}{a^2}\right) = -\frac{3}{a^2}\left(\Hc^2+K\right),\\ \nonumber
G_i{}^j &= -\gamma_i{}^j\left(\frac{K}{a^2}+\frac{2\EX^\tc}{3a}+\frac{\EX^2}{3}\right), \\
&= -\frac{\gamma_i{}^j}{a^2}\left(K+2\Hc^\tc+\Hc^2\right).
\end{align}
For Eq.~\eqref{eq:def:friedpA}, the term $\EX\dEX$ is related to $\EX^2$ in $G_0{}^0$ and $\dRt$ is related to $K/a^2$ in $G_0{}^0$. Hence, one must have
\begin{equation}\label{cond2}
\vert\dEX\vert\ll \vert\EX\vert \rightarrow \left\vert\frac{a\dEX}{\Hc}\right\vert\ll 1,
\end{equation}
and
\begin{equation} \label{cond3}
\left\vert\frac{(D^2+3K)\CS}{K}\right\vert \ll 1.
\end{equation}
In the case where $K=0$, we have to compare $\dRt$ with $G_0{}^0$ and $G_i{}^i$, yielding
\begin{equation} \label{cond31}
\left\vert\frac{D^2\CS}{\Hc^2}\right\vert \ll 1, \quad \left\vert\frac{D^2\CS}{2\Hc^\tc+\Hc^2}\right\vert \ll 1.
\end{equation}

Now we have to establish the conditions on $\dCIS$ and $a^i$, which are null in the background. We will use Eqs.~\eqref{eq:def:friedpCST} and \eqref{eq:def:friedpCSD}, which come from perturbing $G_j{}^i$. These components of the Einstein tensor contain, in the background, $\EX^\tc/a$ and $\EX^2$. The first one originates $\dCIS^\tc/a$ and $\dct_i\pa^i$ in these equations, while the second originates the term $\EX\dCIS$ in Eq.~\eqref{eq:def:friedpCST}. Thus, we obtain the following conditions:
\begin{equation}\label{cond4}
\left\vert\frac{D^2\dCIS}{a\Hc}\right\vert \ll 1, \quad \left\vert\frac{D^2\A}{\Hc^\tc-\Hc^2}\right\vert \ll 1.
\end{equation}
There are no further independent conditions on the geometric perturbations. Note also that each condition was obtained comparing the background and perturbed values of each term of the perturbed Einstein equations. This means that these conditions are stronger than just comparing the perturbed equations with the complete Einstein tensor.

The conditions $$\left\vert\frac{\delta\rho}{\rho}\right\vert \ll 1,\quad\left\vert\frac{\delta{}p}{p}\right\vert\ll 1,$$ when in terms of geometric perturbations using Eqs.~\eqref{eq:def:friedpA} and \eqref{eq:def:friedpCST}, reduce to the conditions already obtained for those quantities. The velocity field potential $\vus$ appears only in the $\delta{}G_i{}^0$ projections, which are null in the background. To deal with this variable we can calculate the second-order correction on the trace of the energy-momentum tensor, which contains a term proportional to $a^{-2}D_i\vus{}D^i\vus$. Such a term also appears when one calculates the normalization of $v_\mu{}v^\mu = -1 + a^{-2}D_i\vus{}D^i\vus$ up to second-order terms, where $v_\mu$ represents the eigenvector of the energy-momentum tensor; we have already considered the other quadratic corrections in this expression much smaller than 1. Therefore, applying the condition $a^{-1}\vert{}D_i\vus\vert \ll 1$ on Eq.~\eqref{eq:def:friedpB}, we obtain
\begin{equation}
\left\vert\frac{aD_i\dEX}{\Hc^2+K-\Hc^\tc}\right\vert\ll1,\quad \left\vert\frac{D_i(D^2+3K)\dCIS}{a(\Hc^2+K-\Hc^\tc)}\right\vert \ll 1.
\end{equation}
These conditions are weaker than Eqs.~\eqref{cond2} and \eqref{cond4} around the bounce phase, first due to the additional factor $\Hc^{-1}\propto\eta$, which goes to zero near the bounce, and also because the spatial derivative brings down higher-order correction terms when acting in the superhorizon solutions (Eq.~\eqref{solform}).

The Weyl tensor $C_{\mu\nu\alpha}{}^\beta$ in a Friedmann metric is null. Therefore, its perturbation will be gauge invariant (see~\cite{Stewart1974}). One could expect to get a gauge invariant condition for the perturbative series by comparing the perturbation of the Weyl tensor with the background Riemann tensor. The non-null components of the perturbed Weyl tensor are given by its electric part $$\delta{}C_{i0}{}^{j0} = a^{-2}\left(D_iD^j-\frac{\gamma_i{}^j}{3}D^2\right)(\AI+\CSI),$$ while the background Ricci tensor components are $$R_{i}{}^{j} = \gamma_i{}^j\left(\frac{2K}{a^2}+\frac{\EX^\tc}{3a}+\frac{\EX^2}{3}\right) = \gamma_i{}^j\frac{2K+\Hc^\tc+2\Hc^2}{a^2}.$$ Comparing these components yields the constraint $$\left\vert\frac{D^2(\AI+\CSI)}{2K+\Hc^\tc+2\Hc^2}\right\vert \ll 1,$$ which is satisfied whenever conditions given in Eqs.~(\ref{cond3}--\ref{cond4}) hold. Hence, the Weyl tensor provides a gauge invariant constraint, which is necessary but not sufficient since it alone does not imply Eqs.~(\ref{cond3}--\ref{cond4}). Additionally, this condition, when applied to the perturbations near the bounce phase is much weaker than that of Eq.~\eqref{cond1}, which is necessary to define the perturbations of the inverse metric.

\subsection{The gauge choice solution}
\label{sec:pert:gauge}

The evolution of the gauge invariant Bardeen potential in the classical GR phase around the bounce is described in Sec.~\ref{sec:genbounce}, where it shown that $\AI$ grows larger than 1 at this phase. In the Newtonian gauge ($\B = 0 = \E$ and consequently $\dCIS = 0$), $\AI = \A$, therefore condition~\eqref{cond1} is not satisfied and linear perturbation theory breaks down in this gauge.

However, choosing a gauge with constant curvature ($\dRt = 0$), we avoid the problem described above.
In this gauge one sets $\psi = \B = 0$ and, in order to completely fix the gauge, we impose $\E(\eta_1)=0$, where $\eta_1$ is some particular convenient choice of conformal time. One has
\begin{equation}\label{eq:def:CIG:dCIS}
\dCIS = -\frac{a\AI}{\Hc}, \quad \A = x^2c_s^2\zz^2\MS, \quad \E = \int^\eta_{\eta_1}\frac{\dd\bar{\eta}\bar{\AI}}{\bar{\Hc}},
\end{equation}
where we used $\AI = \CSI$ ($\dAPD = 0$) and Eq.~\eqref{eq:MS_AI} to obtain the expressions above.

In this gauge the perturbation $\phi$ has a different behavior. At any instant in which a single fluid dominates one has $$\A = \frac{3(1+w)}{2}\MS,$$ where we used Eq.~\eqref{eq:single}. Therefore, in this gauge $\A$ follows the evolution of $\MS$ instead of $\AI$. As we discussed in Sec.~\ref{sec:contracting}, $\MS$ grows in the contracting phase until it attains, near the bounce, an amplitude approximately equal to the constant mode of $\AI$. In the expanding phase, $\MS$ also has a decaying mode, but in this case this mode is always smaller than the constant one. Hence, $\phi \ll 1$ is satisfied in this gauge.

Starting the calculations in the constant curvature gauge, one can see that near the bounce scale the gauge-fixing condition for the Newtonian gauge is not well-defined. In the new gauge the value of $\A$ would change as $3(1+w)/2\MS\rightarrow\AI$. However, as we showed near the bounce $\AI \gg \MS$. This would imply a non valid transformation $$\A_\text{N} = \A_\text{CCG} + \left(\AI - \frac{3(1+w)}{2}\MS\right),$$ where $\A_\text{N}$ represents the metric perturbation in the Newtonian gauge and $\A_\text{CCG}$ in the constant curvature gauge.

The relation between the perturbation $\E$ and $\AI$ is
$$\left\vert\E\right\vert = \left\vert\int^{x_1}_{x}\dd\bar{x}\frac{\bar{x}\bar{\AI}}{\bar{E}^2}\right\vert = \left\vert\left.\frac{x\AI}{E^2}\right\vert_{x_\star}(x_1-x)\right\vert \leq \left\vert\frac{x_b^2\AI(x_b)}{E^2(x_b)}\right\vert,$$ where we used the mean value theorem. Note from Eq.~\eqref{psphibgen} that near the bounce $\AI(x_b) \propto x^{(5+3\wQ)/2}$ and thus $x^2/E^2 \propto x^{-(1+3\wQ)}$; in this gauge $\E \propto x^{3(1-\wQ)/2}$ has the same growth factor as $\A$. With the results above and noting that in this gauge $\psi = \B = 0$, the conditions given in Eq.~\eqref{cond1} are verified. The other conditions given in Eqs.~(\ref{cond2}--\ref{cond4}) can be trivially verified.

\subsection{Gauge choice and the bounce phase}
\label{sec:gauge:bounce}

The discussion above shows that for a variety of models the evolution of the perturbations near a bounce phase is well-behaved in what concerns the perturbative series. However, at the exact moment when the bounce occurs, other problems can arise. The gauge invariant variable $\MS$ is related to the metric perturbations through (Eq.~\ref{eq:def:CIG:dCIS}),
\begin{equation*}
\A = \frac{a^2\kp{}(\rho+p)}{2\Hc^2}\MS.
\end{equation*}
Hence, at the bounce the Hubble function goes to zero and, therefore, the perturbation $\A$ diverges since $\MS$ stays constant at the bounce (see Eq.~\ref{solform}). However, in the synchronous gauge one has
\begin{equation}
\CS = \frac{\Hc}{a}\int\dd\eta \frac{a^3\kp(\rho+p)}{2\Hc^2}\MS,
\end{equation}
and one can show that, using the variable $\mu$ defined as $x = x_b e^{-\mu^2/2}$, one has $\Hc \propto \mu$ near the bounce (this is the case when the bounce is caused by a negative factor in $E^2$). Thus, the integral above is proportional to $\mu\int\dd\mu\mu^{-2}$, and the perturbations are well-behaved at the bounce. Hence, the perturbations are always finite and small, as the constant mode of $\AI$.



\section{Conclusions}
\label{conclusions}

We have shown in this paper that, for adiabatic perturbations, the Bardeen 
potential in the contracting phase of bouncing models can generally become 
very large, but this fact does not invalidate linear perturbation theory 
around the bounce. We established necessary conditions for the validity of 
linear perturbation theory on Friedmann backgrounds, and we have shown
that there is a gauge choice, for a large class of bouncing models, 
where these conditions are satisfied. In fact, there are some gauges that
are ill-defined close to the bounce because the gauge transformations
relating them to some well-behaved gauge are not valid.

In conclusion, the program of describing the evolution
of linear primordial perturbation in bouncing models is 
well-defined. However, one must take care with the gauge which 
will be chosen while performing calculations, since some
gauge fixing conditions are not well-behaved in these scenarios.
In general, the gauge invariant approach is more appropriated.
It does not depend on any gauge fixing condition. However this
approach alone is not enough to evaluate the validity of the
linear approximation. Therefore, one must always check if 
there is a gauge in which the perturbation series is valid.

\begin{acknowledgments}
This work was supported by Conselho Nacional de Desenvolvimento Cient\'{i}fico e Tecnol\'ogico (CNPq) -- Brasil.
\end{acknowledgments}

\appendix*

\section{Asymptotic Series Approximation}
\label{app:intaprox}
The integral \eqref{intgen1} can be rewritten as
\begin{equation}
\label{intgen11}
\int^{x_c}_0\frac{\dd{}x}{E\zz^2} = \int^{\nu_c}_{-\infty}\dd\nu g(\nu)e^{s\nu},
\end{equation}
where we have defined $\nu = \ln(x)$, $s\equiv 3(1-\wQ)/2$, $\nu_c = \ln(x_c)$, and $g(\nu) \equiv 1/(x^{(1-3\wQ)/2}E\zz^2)$. This divides the integrand in the dominant term $e^{s\nu} = x^{3(1-\wQ)/2}$ and the controlled function $g(\nu)$. Using Eq.~\eqref{approxz2}, it can be written as
\begin{equation}
\label{ggen}
g(\nu) = \frac{1}{x^{(1-3\wQ)/2}E\zz^2} = \frac{2 c_s^2 x^{3(1+\wQ)/2}}{3E(1+p/\rho)}.
\end{equation}
Hence, $$\lim_{\nu\rightarrow \nu_c}g(\nu) = \OO{1}, \quad \lim_{\nu\rightarrow-\infty}g(\nu) = 0.$$

The first limit comes from the fact that $E \propto x^{3(1+\wQ)/2}$ for $x\rightarrow x_c$. This could also be seen in the situation where the matter content, besides the fluid with $p_q/\rho_q = \wQ$, is given by a collection of $n$ other constant $w_i$ fluids, yielding
\begin{equation*}
g(\nu) = \frac{2c_s^2\Omega_{w_q0}^{-1/2}}{3(1+p/\rho)}\left[\sum_{i=1}^n{}\frac{\Omega_{w_i0}}{\Omega_{w_q0}}x^{-3(\wQ-w_i)} + 1\right]^{-1/2}.
\end{equation*}
As $\wQ > w_i$, when in the domain in which $x \gg 1$, $g(\nu)$ is dominated by a constant value of order one.

The second limit corresponds to $x\rightarrow 0$ or $a\rightarrow\infty$, and one expects that the universe was dominated by the fluid with $w_1$ since we ordered the fluids imposing $w_i < w_{i+1}$, giving $g(\nu) \propto x^{3(\wQ-w_1)/2} \rightarrow 0$.

With these results, we can integrate Eq.~\eqref{intgen11} by parts to obtain
\begin{equation}
\int^{x_c}_0\frac{\dd x}{E\zz^2} = \frac{e^{s\nu_c}}{s}g(\nu_c) - \int_{-\infty}^{\nu_c}\dd\nu\frac{e^{s\nu}}{s}\frac{\del{}g(\nu)}{\del\nu}.
\end{equation}
Note that $$\lim_{\nu\rightarrow \nu_c}x^r\frac{\del{}g}{\del\nu} = \OO{1},\quad \lim_{\nu\rightarrow -\infty}x^r\frac{\del{}g}{\del\nu} = 0,$$ where $r={3(\wQ-w_n)}$. Thus, we can again split the integrand in a controlled function $g_1 \equiv x^r\del{}g/\del\nu$ times $e^{(s-r)\nu}$, and integrate by parts obtaining
\begin{align*}
&\int^{x_c}_0\frac{\dd x}{E\zz^2} = \\
&\frac{e^{s\nu_c}}{s}g(\nu_c) - \frac{e^{s\nu_c}}{s(s-r)}\left.\frac{\del{}g}{\del\nu}\right\vert_{\nu_c} + \int_{-\infty}^{\nu_c}\dd\nu{}e^{(s-r)\nu}\frac{\del{}g_1}{\del\nu}.
\end{align*}
The magnitude of the second term has an additional factor of $x_c^{-r}$ when compared with the first. Repeating the process of factoring the largest growing term and integrating by parts, we obtain an asymptotic series approximation for this integral. For a complete discussion about this method of approximating integrals, see~\cite{Wong2001}. 

As we have shown for the first term, each subsequent term will by multiplied by an additional factor of $x_c$ to a negative power. Therefore, to estimate the order of magnitude of the integral, it is sufficient to keep only the first term,
\begin{equation}\label{eq:app:intfapprox}
\int^{x_c}_0\frac{\dd{}x}{E\zz^2} \approx \frac{2}{3(1-\wQ)}\frac{x_c}{E(x_c)\zz(x_c)^2}.
\end{equation}
It is worth noting that the result above is exact if the matter content consists in just a single fluid with constant equation of state.


\end{document}